\documentclass[twocolumn,prb,showpacs,superscriptaddress,preprintnumbers]{revtex4}
\usepackage{amsmath}

\usepackage{graphicx}
\usepackage{dcolumn}
\usepackage{bm}

\begin{document}

\title{Anomalous Quantum Interference Induced by Landau-Zener Transitions
in a Strongly Driven rf-SQUID Qubit}
\author{Yiwen Wang}
\affiliation{National Laboratory of Solid State Microstructures and Department of
Physics, Nanjing University, Nanjing 210093, China}
\author{Shanhua Cong}
\affiliation{Department of Electronic Science and Engineering,Research Institute of
Superconductor Electronics, Nanjing University, Nanjing 210093, China}
\author{Xueda Wen}
\affiliation{National Laboratory of Solid State Microstructures and Department of
Physics, Nanjing University, Nanjing 210093, China}
\author{Cheng Pan}
\affiliation{National Laboratory of Solid State Microstructures and Department of
Physics, Nanjing University, Nanjing 210093, China}
\author{Guozhu Sun}
\email{gzsun@nju.edu.cn}
\affiliation{Department of Electronic Science and Engineering,Research Institute of
Superconductor Electronics, Nanjing University, Nanjing 210093, China}
\author{Jian Chen}
\affiliation{Department of Electronic Science and Engineering,Research Institute of
Superconductor Electronics, Nanjing University, Nanjing 210093, China}
\author{Lin Kang}
\affiliation{Department of Electronic Science and Engineering,Research Institute of
Superconductor Electronics, Nanjing University, Nanjing 210093, China}
\author{Weiwei Xu}
\affiliation{Department of Electronic Science and Engineering,Research Institute of
Superconductor Electronics, Nanjing University, Nanjing 210093, China}
\author{Yang Yu}
\email{yuyang@nju.edu.cn}
\affiliation{National Laboratory of Solid State Microstructures and Department of
Physics, Nanjing University, Nanjing 210093, China}
\author{Peiheng Wu}
\affiliation{Department of Electronic Science and Engineering,Research Institute of
Superconductor Electronics, Nanjing University, Nanjing 210093, China}

\begin{abstract}
We irradiated an rf-SQUID qubit with large-amplitude and high frequency
electromagnetic field. Population transitions between macroscopic
distinctive quantum states due to Landau-Zener transitions at energy-level
avoided crossings were observed. The qubit population on the excited states as a
function of flux detuning and microwave power exhibits
interference patterns. Some novel features are found in the interference and
a model based on rate equations can well address the features.
\end{abstract}

\pacs{74.50.+r, 85.25.Cp}
\maketitle

Coherent quantum dynamics of superconducting Josephson qubits driven by weak
external field have been extensively investigated in experiments. \cite%
{Nakamura,Vion,Yu,Chio,Satio,Lisenfeld} These experiments used Rabi
oscillation to probe the macroscopic quantum coherence of such systems in
the time domain, where the driving frequency equals to the energy-level
separation. On the other hand, quantum coherence can also be investigated
with a large-amplitude field, which can drive the qubit throughout its
energy-level spectrum. In this case, Landau-Zener (LZ) transition process
occurs around each level-crossing position with a finite probability while
the quantum evolution is adiabatic away from the level-crossing positions.%
\cite{Shytov,Shev} Repetition of LZ transitions gives rise to Stuckelberg or
Ramsey-type oscillations \cite{Stuck,Ramsey} in analogy to Mach-Zehnder (MZ) interferometer. The MZ-type interference is a unique quantum coherence signature complementary to Rabi
oscillation.

Recently, coherent dynamics of superconducting qubits in the strongly driven
regime dominated by LZ transitions were extensively studied, especially in
two level systems, providing an alternative method to manipulate and
characterize the qubit.\cite{Sill,Izma,Shev2,Oliver,Berns} A recent work
makes a step forward \cite{Berns2} : the superconducting flux qubit as a
multi-level system was driven through several avoided crossings between
energy levels. The qubit population under large-amplitude fields exhibited a
series of diamond-like interference patterns in the space parameterized by
static flux detuning and microwave amplitude. Spectroscopic information can
be obtained from the field amplitude dependence of the qubit population.
\cite{Rudner} That experiment was done with a superconducting
persistent-current qubit, which has relatively a small loop size and a long
coherence time. The applied microwave frequency was much smaller than the
energy-level spacing between discrete energy levels and therefore the
resonant peaks/dips tend to merge into a continuous band.

Comparing with a superconducting persistent-current qubit, an rf-SQUID qubit
has a much larger geometric loop size (about 100 times larger in loop area)
which usually leads to a very short coherence time ($\sim $1 ns). In order
to observe LZ interference in such a large system, the time interval between
subsequent LZ tunneling events should be less than the relevant coherence
time, indicating a high-frequency field is required. In a previous work,
\cite{Guozhu} we observed the quantum interference fringes in an rf-SQUID
qubit due to LZ transitions at one energy-level avoided crossing. In this
paper, we report the observation of simultaneous presence of two-set of
interference fringes associating with LZ transitions at two nearby avoided
crossings. In addition, the qubit population can be modulated by the field
amplitude at certain static flux bias points. We developed a model based on
rate equations and addressed these features very well.

Our sample was fabricated with Nb/AlO$_{x}$/Nb trilayer on an oxidized Si
wafer by using the standard photolithography method. As shown in Fig. 1(a)
and 1(b), the rf-SQUID qubit is essentially a superconducting loop with a
second order gradiometric configuration interrupted by a small dc-SQUID.
When an external magnetic flux $\Phi _{f}^{q}$ colse to $\Phi _{0}/2$, where
$\Phi _{0}$ is the flux quantum, is applied to the qubit loop, the system
potential takes a double well shape. The flux states in different wells,
serving as the qubit states, correspond to macroscopic circulating currents
flowing in opposite directions. Two on-chip current bias lines are used to
control the magnetic flux applied to the qubit loop ($\Phi _{f}^{q}$) and
the small dc-SQUID loop ($\Phi _{f}^{CJJ}$) respectively. $\Phi _{f}^{q}$
determines the asymmetry of the potential and $\Phi _{f}^{CJJ}$ modulates
the potential barrier height. By adjusting $\Phi _{f}^{q}$ and $\Phi
_{f}^{CJJ}$, the shape of the system potential can be fully controlled $in$ $%
situ$. An additional on-chip current bias line is used to supply microwave (MW)
pulses. The qubit flux states can be read out by a hysteretic dc-SQUID
inductively coupled to the qubit loop. We can easily bias the readout
dc-SQUID at its maximum sensitivity region by applying an external magnetic
flux ($\Phi _{f}^{dc}$).

\begin{figure}[tbp]
\centering
\includegraphics[width=3.2375in]{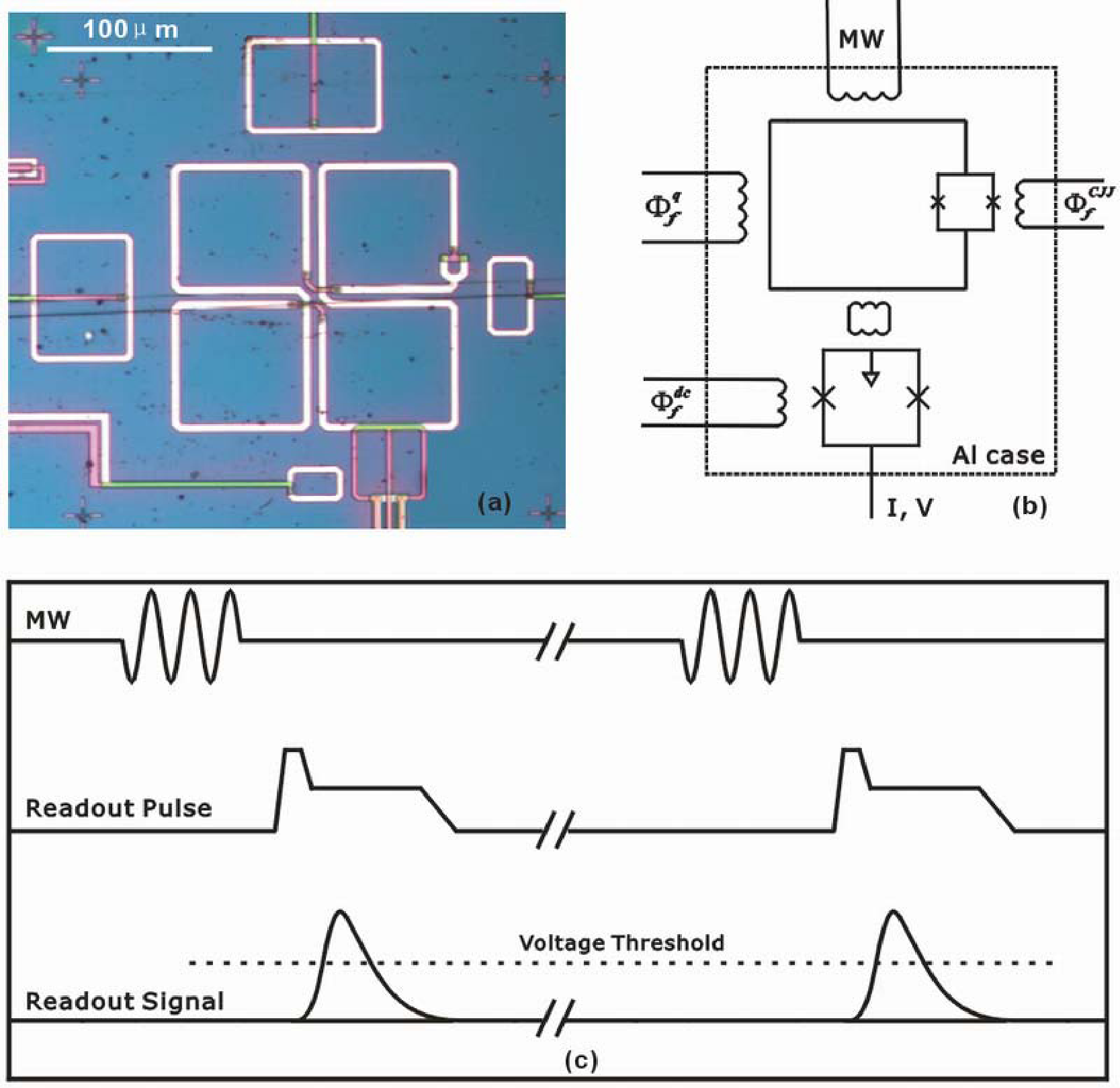}
\caption{(color online) (a) Optical micrograph of an rf-SQUID qubit (big
loop interrupted by a small dc-SQUID with two Josephson junctions). (b)
Schematic of the on-chip circuits (crosses represent the Josephson
junctions). (c) The measurement time sequence . }
\end{figure}

\begin{figure*}[tbp]
%[htb]
\centering
\includegraphics[width=5.5in]{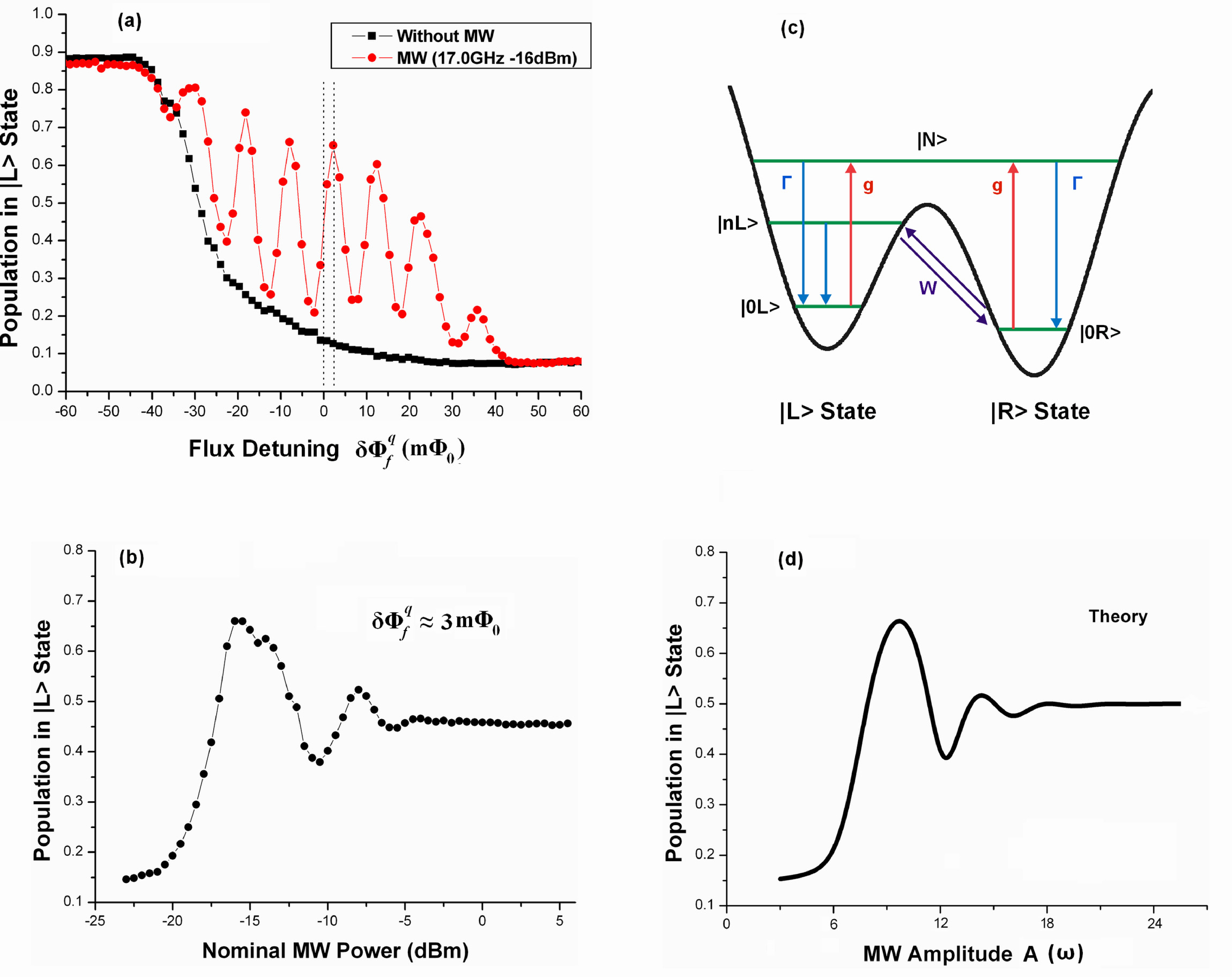}
\caption{(color online) (a) Normalized qubit population in $|L\rangle $
state as a function of static flux detuning $\Phi _{f}^{q}$. Four peaks and
dips are induced by a 10 $\protect\mu $s microwave pulse at 17.0 GHz and
-16dBm. As a guide to eyes, the left dotted line marked the midpoint of the
peak/dip positions. Without microwave, the qubit transition step center is
displaced from this midpoint. The right dotted line corresponds to static
flux detuning $\protect\delta \Phi _{f}^{q}$ = 3m$\Phi _{0}$ (b) Normalized
qubit population in $|L\rangle $ state versus the nominal microwave power at
a fixed flux detuning 3 m$\Phi _{0}$. (c) Schematic of the relevant
transition processes involved in Fig. 2 (b). $|0R\rangle $ and $|nL\rangle $
are the ground states of right well and left well respectively. $|N\rangle $
represents a higher energy level which is not localized in either potential
well. The red (blue) lines show the excitation (relaxation) path. The purple
lines represent the LZ transitions between $|0R\rangle $ and $|nL\rangle $.
(d) Calculated qubit population in $|L\rangle $ state as a function of
microwave amplitude. The relevant parameters used in our theoretical model
are: $\omega/2\pi $ = 17 GHz, $m$ = 8, $\Delta/2\pi $ = 7 MHz, $\Gamma _{2}/2\pi$ = 2 GHz, $\protect\gamma/2\pi $ = $%
\Gamma/2\pi $ = 2 GHz, $\Gamma_{10}/2\pi=0.6$ KHz, $\Gamma _{01}/2\pi$ = 0.1 KHz, a$/2\pi$ = 5
Hz, b = 1.4.}
\end{figure*}

The sample was placed on a chip carrier enclosed in a superconducting
aluminum sample cell and the experiments were performed in a dilution
refrigerator at a base temperature about 20 mK. The device was magnetically
shielded and all electrical lines were carefully filtered and equipped with
proper cold attenuators to minimize circuit noise (see Ref.\cite{Guozhu2}
for detailed description).

In our experiment, we biased $\Phi _{f}^{CJJ}$ at a certain value such that
the critical current of the small dc-SQUID is close to its maximum.
Therefore the barrier between two potential well maintained at a relatively
high value. There are several flux states below the top of the barrier in
each potential well. The inter-well flux state transition rate between the
ground states of two potential wells is very small at flux biases close to $%
\Phi _{0}/2$. Since the intra-well relaxation time is very short, the state
of the qubit freezes quickly after turning off microwave irradiation. The
time sequence of measurement is shown in Fig. 1(c). For each measurement
cycle, a 10 $\mu $s microwave pulse is applied to the qubit loop to induce
sinusoidal excursions through the energy levels about a static flux detuning
$\delta \Phi _{f}^{q}$ = $\Phi _{f}^{q}$ - $\Phi _{0}/2$. After a delay time
about 1 $\mu $s, the readout is performed by driving the dc-SQUID with a
current pulse comprising a 15 ns sampling current followed by a 15 $\mu $s
holding current just above the retrapping current of the dc-SQUID. By properly
adjusting the magnitude of the readout current pulse, the dc-SQUID either
switches to finite-voltage state or stays at zero voltage state
corresponding to the qubit states being in left well $|L\rangle $ or in
right well $|R\rangle $. The measurement repetition frequency is 5 KHz. By
repeating the trials for 4$\times $10$^{4}$ times, we obtained the average
switching probability, representing the population in $|L\rangle $ state. By
changing the flux detuning and microwave power step by step, we then
measured the dependence of qubit population in $|L\rangle $ state on flux
detuning and microwave power.

\begin{figure}[tbp]
\centering
\includegraphics[width=3.2375in]{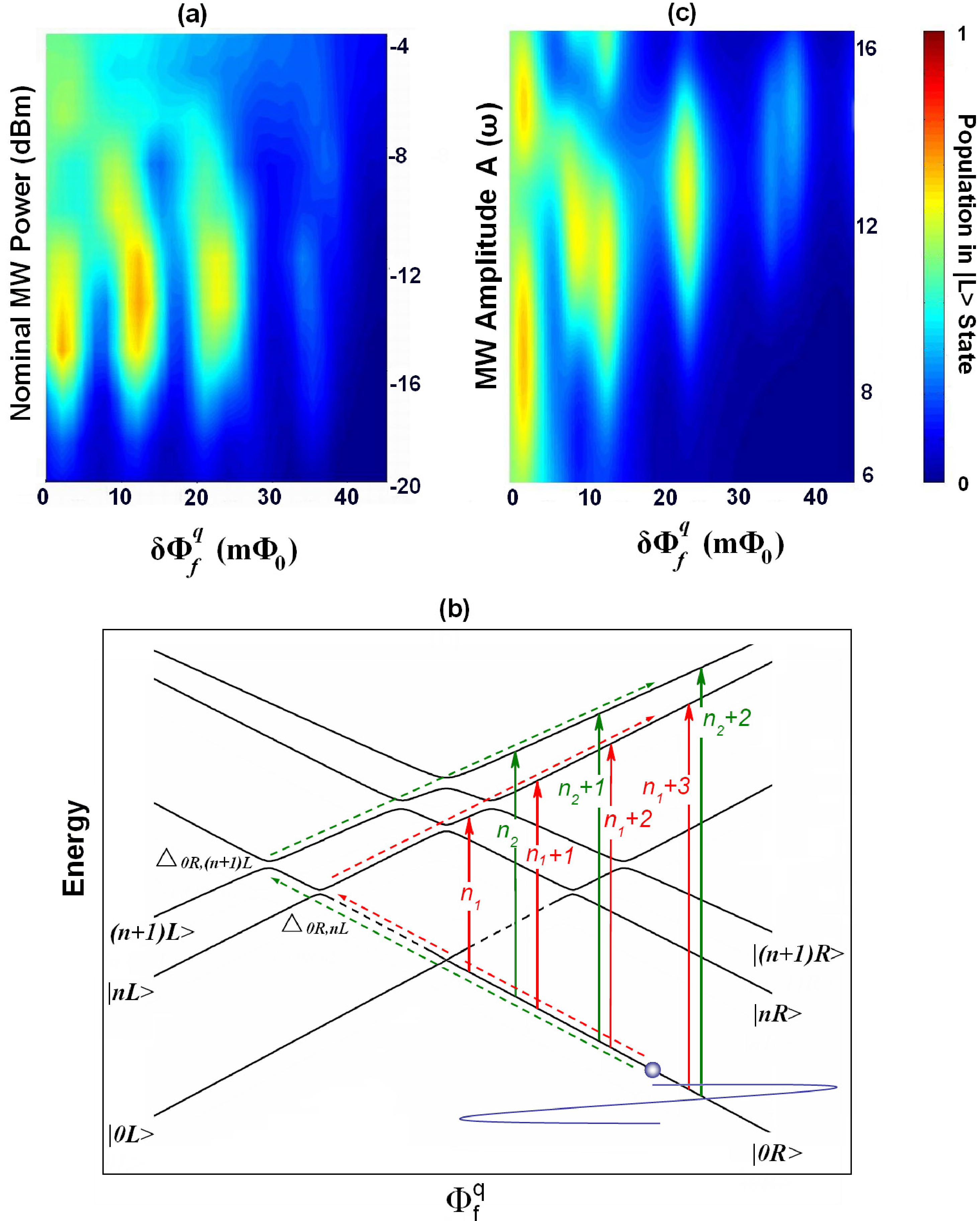}
\caption{(color online) (a) Dependence of qubit population in $|L\rangle $
state on static flux detuning $\protect\delta \Phi _{f}^{q}$ and nominal
microwave power, with microwave frequency at 17.0 GHz. Part of two
overlapped interference sets can be observed and population inversion is
clearly demonstrated. (b) Energy-level diagram illustrating the multi-photon
transition processes. The qubit is initially in $|0R\rangle $, the ground
state in right well. Microwave harmonically drives the qubit through the
energy levels that are approximately linear in flux detuning and LZ
transitions occur at the avoided level crossings $\Delta _{0R,nL}$ and $%
\Delta _{0R,(n+1)L}$ (another two level crossings locating at positive flux
detuning are not shown in this figure). The red and green path, with
different energy-level slope, represent the main processes to generate the
first and second interference set respectively. (b) Numerical calculation
of qubit population in $|L\rangle $ state versus flux detuning and microwave
amplitude. The calculated interference patterns show good agreement with the experimental
observations (Fig. 3(a)). }
\end{figure}

Fig. 2(a) shows the resonant peaks and dips when applying a microwave with
frequency 17.0 GHz and nominal power -16 dBm. The peaks (dips) reflect the
mcrowave induced excitation from flux state $|R\rangle $ to $|L\rangle $ ($%
|L\rangle $ to $|R\rangle $). The width of the resonant peaks/dips is
generally on the order of several m$\Phi _{0}$ and the decoherence
time of our system is estimated to be less than 1 ns.
The strong population inversion is evident, suggesting that higher energy
levels other than the first two lowest states are involved. \cite{Guozhu}
Without microwave the qubit step center is shifted away from the symmetry
position of the energy spectrum (marked by the left dotted line in Fig.
2(a)) due to the presence of the circulating current in the readout
dc-SQUID. \cite{Chio} Then we fixed the flux bias and swept microwave power. Fig. 2(b) shows the qubit
population in $|L\rangle $ state as a function of the nominal microwave
power at a static flux detuning $\delta \Phi _{f}^{q}$ = 3 m$\Phi _{0}$
(marked by the right dotted line in Fig. 2(a)). The population exhibits a
damped oscillation around 0.5. For two level systems, a good Bessel function
dependence of the population on microwave amplitude was observed in a
superconducting persistent current qubit and the modulation period is
proportional to the ratio of microwave amplitude and the microwave
frequency. \cite{Oliver} Therefore, in our experiments a large microwave
power is required to observe the periodic behavior since the microwave
frequency is very large. However, for a very high microwave power, the LZ
transitions are not the dominant processes anymore. To make a more
quantitative understanding, we hereby propose a simple model, which is
illustrated in Fig. 2(c). The qubit is initialized in $|0R\rangle $, the
ground state of right well. When the microwave amplitude is not very large
(i.e., the double well potential always exits), the LZ transitions (with
rate $W$) between $|0R\rangle $ and $|nL\rangle $ (the $n$th excited state
of left well), and the fast intra-well relaxation (with rate $\gamma $) from
$|nL\rangle $ to $|0L\rangle $ dominate the dynamics of the qubit
population. These processes cause qubit population inversion and lead to the
non-monotonic dependence of qubit population on microwave power. When the
microwave amplitude goes up, it is much easier for the the qubits to be
directly excited from the ground state to a higher energy level $|N\rangle $
(with rate $g$), which is above the top of the barrier and has a wavefunction spanning both potential wells. The
qubit population in $|N\rangle $ tend to fast relax back to the ground
states ($|0L\rangle $ and $|0R\rangle $) of both wells with an approximately
equal relaxation rate $\Gamma $ and therefore this process equalizes the
qubit population in the two wells. Taking all transition processes into
consideration, one can obtain the rate equations to describe the qubit level
occupations $p_{i}(i=0,1,2,3$, corresponding to $|0R\rangle $, $|0L\rangle $%
, $|nL\rangle $ and $|N\rangle $ respectively) as:
\begin{align}
& \dot{p}_{0}=-(W+g+\Gamma _{01})p_{0}+\Gamma _{10}p_{1}+Wp_{2}+\Gamma p_{3},
\nonumber \\
& \dot{p}_{1}=\Gamma _{01}p_{0}-(g+\Gamma _{10})p_{1}+\gamma p_{2}+\Gamma
p_{3},  \nonumber \\
& \dot{p}_{2}=Wp_{0}-(W+\gamma )p_{2},  \nonumber \\
& p_{0}+p_{1}+p_{2}+p_{3}=1
\end{align}

Under the assumption of classical noise and using perturbation theory, the
LZ transition rate between $|0R\rangle $ and $|nL\rangle $ takes the form%
\cite{Rudner}:

\begin{equation}
W(\epsilon,x)=\frac{\Delta^{{2}}}{2}\sum_{m}\frac{\Gamma_{2}J^{2}_{m}(x)}{%
(\epsilon-m\omega)^{2}+\Gamma^{2}_{2}}
\end{equation}
where $\Delta$ is the avoided crossing between $\vert 0R \rangle$ and $\vert
nL \rangle$, $\epsilon$ is the dc energy detuning from the avoided crossing $%
\Delta$ and $\Gamma_{2}=1/T_{2}$ is the system dephasing rate. $J_{m}(x)$
are Bessel functions of the first kind with the argument $x=A/\omega$, where
$A$ and $\omega$ are field amplitude and frequency respectively. For a $m$%
-photon transition process, $\epsilon=m\omega$ when on resonance, the LZ
transition rate approximately reduces to:
\begin{equation}
W(\epsilon,x)=\frac{\Delta^{{2}}}{2}\frac{J^{2}_{m}(x)}{\Gamma_{2}}
\end{equation}
The transition rate from the ground state to $|N\rangle $ is sensitive to the barrier height during the microwave driving process. We simply assumed the transition rate have an exponential dependence on field amplitude $A$, i.e., $g$ = $a$$%
e^{bA}$, where $a$, $b$ are two fitting parameters. $\Gamma _{10}$ ($\Gamma
_{01}$) is the slow inter-well relaxation rate between $\vert 0L \rangle$ and $\vert
0R \rangle$. In the stationary case, $\dot{p}_{0}=\dot{p}_{1}=\dot{p}_{2}=%
\dot{p}_{3}=0$, Eq.(1) can be solved numerically with appropriate model parameters. Fig. 2(d) shows the calculated qubit population of state $%
|L\rangle $ as a function of the microwave amplitude. The theoretical curve
clearly exhibits the main features of the experimental results, indicating
our model captured the underlying physics of the system.

Fig. 3(a) shows the contour plot of the qubit population in state $|L\rangle
$ versus the positive static flux detuning $\delta \Phi _{f}^{q}$ and the
nominal microwave power with a fixed microwave frequency at 17.0 GHz.
According to the previous works\cite{Oliver}, we expect that the resonant
peaks are equally spaced on the flux axis and the distance between the adjacent
peaks corresponds to the microwave frequency. With the amplitude increasing,
the resonant peaks emerge one by one, starting from the flux close to $\Phi
_{0}/2.$ The flux position for each peak almost does not change while the
population follows the Bessel functions of the first kind. However, here we
observed two sets of the resonant peaks. The first set includes four
resonant peaks, locating at around flux detuning 3 m$\Phi _{0}$, 12 m$\Phi
_{0}$, 23 m$\Phi _{0}$, and 37 m$\Phi _{0},$ respectively. The second set
lies in the higher microwave power region and includes three resonant peaks,
locating at around flux detuning 10 m$\Phi _{0}$, 22 m$\Phi _{0}$, and 38 m$%
\Phi _{0},$ respectively. We can group them into two sets because as
microwave frequency varies, their flux bias positions move with two different
energy-spectrum slopes, indicating two different transition paths are present. The two interference sets have overlap along microwave power axis. In addition, the
distance of the resonant peak increases with the flux detuning. Therefore,
the interference pattern looks a little complicated and is different from the regular
patterns induced by Landau-Zener transition\cite%
{Sill,Izma,Shev2,Oliver,Berns,Guozhu}. In order to explain the interference
we have to deal with the multi-level system. Since the barrier of the
rf-SQUID is very high, the Landau-Zener transition rate is very low for the
lower energy states. We thereby consider the avoided crossings close to the
top of the barrier, which is qualitatively illustrated in Fig. 3(b). The
qubit is initially in state $|0R\rangle $ at a certain static flux bias $%
\Phi _{f}^{q}$. The first set is associated with LZ transitions at level
crossing $\Delta _{0R,nL}$ (also $\Delta _{0L,nR}$, not shown in Fig. 3(a)
for simplicity) and the red path represents the main transition processes
for generating the first set. The second set is associated with LZ
transitions at level crossing $\Delta _{0R,(n+1)L}$ (also $\Delta
_{0L,(n+1)R}$) and the green path represents the main transition processes
for generating the second set. The fours peaks in the first set
correspond to $n_{1}$-photon and up to ($n_{1}$+3)-photon transitions and
the three peaks in the second set correspond to $n_{2}$-photon and up to ($%
n_{2}$+2)-photon transitions. The slope of the green path is smaller than
that of the red path. Therefore the flux spacing of the resonant peaks of
the second set is larger than that of the first set. On the other hand, the
slopes of the energy levels close to the top of the barrier are flattened,
leading to the increase of the distance of the resonant peaks. Moreover, the
most notable feature is the overlap of these two sets of the resonant peaks,
which can also be understood from Fig. 3(b). For a high-frequency microwave,
the peak region of the interference fringe of the first set can cover a wide
range of microwave power. Therefore, during the modulation of the first set,
the second level crossing can be easily reached and then the second
interference set starts to show up.  This feature is different from the
observation of well-resolved individual interference diamonds when using a
relatively low microwave frequency.

We can quantitatively compare the experimental results with the theory by
performing numerical calculation based on the rate equation approach introduced
in Ref. \cite{Xueda}. In our calculation, the six energy levels shown in
Fig. 3(b) are taken into account. For convenience, we use $|0\rangle $, $%
|1\rangle $, $|2\rangle $, $|3\rangle $, $|4\rangle $, $|5\rangle $ to
denote flux states $|0R\rangle $, $|0L\rangle $, $|nR\rangle $, $|nL\rangle $%
, $|(n+1)R\rangle $, $|(n+1)L\rangle $ respectively. $\Delta _{0R,nL}$ and $%
\Delta _{0R,(n+1)L}$ are also denoted by $\Delta _{1}$ and $\Delta _{2}$.
Then we can write down the rate equations to describe the qubit level
occupations $p_{i}(i=0,1,2,3,4,5)$ as
\begin{align}
& \dot{p}_{0}=-(W_{03}+W_{05}+\Gamma _{01})p_{0}+\Gamma _{10}p_{1}+\Gamma
(p_{2}+p_{4})  \nonumber \\
& \phantom{\dot{p}_{0}=}+W_{30}p_{3}+W_{50}p_{5},  \nonumber \\
& \dot{p}_{1}=\Gamma _{01}p_{0}-(W_{12}+W_{14}+\Gamma _{10})p_{1}+\Gamma
(p_{3}+p_{5})  \nonumber \\
& \phantom{\dot{p}_{1}=}+W_{21}p_{2}+W_{41}p_{4},  \nonumber \\
& \dot{p}_{2}=W_{12}p_{1}-(W_{21}+\Gamma )p_{2},  \nonumber \\
& \dot{p}_{3}=W_{03}p_{0}-(W_{30}+\Gamma )p_{3},  \nonumber \\
& \dot{p}_{4}=W_{14}p_{1}-(W_{41}+\Gamma )p_{4},  \nonumber \\
& p_{0}+p_{1}+p_{2}+p_{3}+p_{4}+p_{5}=1
\end{align}%
where $W_{ij}$ are the LZ transitions between states $|i\rangle $ and $%
|j\rangle $, which can be explicitly written as:
\begin{equation}
W_{03}=W_{30}=W_{12}=W_{21}=\frac{\Delta _{1}^{{2}}}{2}\sum_{m}\frac{\Gamma
_{2}J_{m}^{2}(x)}{(\epsilon _{1}-m\omega )^{2}+\Gamma _{2}^{2}},  \nonumber
\\
\end{equation}%
\begin{equation}
W_{05}=W_{50}=W_{14}=W_{41}=\frac{\Delta _{2}^{{2}}}{2}\sum_{m}\frac{\Gamma
_{2}J_{m}^{2}(x)}{(\epsilon _{2}-m\omega )^{2}+\Gamma _{2}^{2}}
\end{equation}%
where $\epsilon _{1}$ and $\epsilon _{2}$ are the dc energy detuning from
the avoided crossing $\Delta _{1}$ and $\Delta _{2}$ respectively. $\Gamma $
is the fast intra-well relaxation rate from the excited state to the ground
state and $\Gamma _{10}$ ($\Gamma _{01}$) is the slow inter-well relaxation
rate between $|1\rangle $ and $|0\rangle $. In the stationary case, $\dot{p}%
_{0}=\dot{p}_{1}=\dot{p}_{2}=\dot{p}_{3}=\dot{p}_{4}=\dot{p}_{5}=0$. In our
calculation, $\omega/2\pi $ = 17 GHz, $n$ = 4, $n_{1}$ = 8, $n_{2}$ = 10, $\Delta
_{0R,4L}/2\pi$ = $\Delta _{1}/2\pi$ = 7 MHz, $\Delta _{0R,5L}/2\pi$ = $\Delta _{2}/2\pi$ = 13
MHz. As shown in Fig. 3(b), these model parameters describes the
energy-level structure, which can be determined from the parameters: qubit
loop inductance $L$ $\approx $ 1.3 nH, junction capacitance $C$ $\approx $
35 fF and junction critical current $I_{c}$ $\approx $ 610 nA. Other
relevant parameters in calculation are: $\Gamma _{2}/2\pi$ = 2 GHz, $\Gamma/2\pi $ = 2
GHz, and $\Gamma _{10}$ = 0.5 KHz$\times $$e^{\beta \epsilon _{10}}$ ($%
\epsilon _{10}$ is the energy difference between $|1\rangle $ and $|0\rangle
$). By using the above parameters and taking into account the different
energy-level slope for different transition paths, one can numerically solve
Eq.(4) in the stationary case and obtains Fig. 3(c), showing the qubit population in $|L\rangle $
state as a function of the microwave amplitude and static flux detuning. The
agreement between experimental results and numerical calculation is
remarkable.

It has been known that the quantum interference patterns in a
superconducting flux qubit is dominated by the microwave frequency and qubit
decoherence time\cite{Xueda}. Due to the larger size, the rf-SQUID qubit
usually has much shorter decoherence time than that of the superconducting
persistent current qubit. Therefore, in order to resolve interference
pattern one has to apply microwave with large frequency. Since the response
of the microwave is proportional to $A/\omega ,$ a large microwave power
is subsequently required. The large power and frequency will lead excitation
to many energy levels. Although the contribution from high levels
complicates the interference pattern, we can still quantitatively understand
the interference using LZ transition. For more theoretical discussions on the interference features with large-amplitude and high-frequency field, one may refer to our recent note.\cite{Xueda2}

In conclusion, we observed the quantum interference fringes in a
superconducting rf-SQUID qubit driven by large-amplitude and high-frequency
microwave field. The interference pattern exhibits two sets of overlapped
resonant peaks with well-resolved multi-photon transitions. The two
interference sets correspond to two different transition paths and each path
is associated with LZ transitions at a particular level-crossing position.
By considering the energy levels close and above the top of the energy
barrier between the potential wells, we numerically calculated interference
patterns. The agreement between the simulation and experimental results
reflects the multi-level feature of the rf-SQUID qubit.

This work was partially supported by NSFC (10674062, 10704034, 10725415),
the State Key Program for Basic Research of China (2006CB921801).


\begin{thebibliography}{0}
\expandafter\ifx\csname natexlab\endcsname\relax\def\natexlab#1{#1}\fi
\expandafter\ifx\csname bibnamefont\endcsname\relax
  \def\bibnamefont#1{#1}\fi
\expandafter\ifx\csname bibfnamefont\endcsname\relax
  \def\bibfnamefont#1{#1}\fi
\expandafter\ifx\csname citenamefont\endcsname\relax
  \def\citenamefont#1{#1}\fi
\expandafter\ifx\csname url\endcsname\relax
  \def\url#1{\texttt{#1}}\fi
\expandafter\ifx\csname urlprefix\endcsname\relax\def\urlprefix{URL }\fi
\providecommand{\bibinfo}[2]{#2}
\providecommand{\eprint}[2][]{\url{#2}}

\end{thebibliography}


\begin{thebibliography}{99}
\bibitem{Nakamura} Y. Nakamura, Yu. A. Pashkin, and J. S. Tsai, Phys. Rev.
Lett. \textbf{87}, 246601 (2001).

\bibitem{Vion} D. Vion, A. Aassime, A. Cottet, P. Joyez, H. Pothier, C.
Urbina, D. Esteve, and M. H. Devoret, Science \textbf{296}, 886 (2002).

\bibitem{Yu} Y. Yu, S. Han, X. Chu, S. Chu, and Z. Wang, Science \textbf{296}%
, 889 (2002).

\bibitem{Chio} I. Chiorescu, Y. Nakamura, C. J. P. M. Harmans, and J. E.
Mooij, Science \textbf{299}, 1869 (2003).

\bibitem{Satio} S. Satio, T. Meno, M. Ueda, H. Tanaka, K. Semba, and H.
Takayanagi, Phys. Rev. Lett. \textbf{96}, 107001 (2006).

\bibitem{Lisenfeld} J. Lisenfeld, A. Lukashenko, M. Ansmann, J. M. Martinis,
and A. V. Ustinov, Phys. Rev. Lett. \textbf{99}, 170504 (2007).

\bibitem{Shytov} A. V. Shytov, D. A. Ivanov, and M. V. Feigel'man, Eur.
Rhys. J. B. \textbf{36}, 263 (2003).

\bibitem{Shev} S. N. Shevchenko, A. S. Kiyko, A. N. Omelyanchouk, and W.
Krech, Low Temp. Phys. \textbf{31}, 569 (2005).

\bibitem{Stuck} E. C. G. Stuckelberg, Helv. Phys. Acta \textbf{5}, 369
(1932).

\bibitem{Ramsey} N. F. Ramsey, Phys. Rev. \textbf{76}, 996 (1949).

\bibitem{Sill} M. Sillanpaa, T. Lehtinen, A. Paila, Y. Makhlin, and P.
Hakonen, Phys. Rev. Lett. \textbf{96}, 187002 (2006).

\bibitem{Izma} A. Izmalkov, S. H. W. can der Ploeg, S. N. Shevchenko, M.
Grajcar, E. Il'ichev, U. Hubner, A. N. Omelyanchouk, and H.-G.Meyer, Phys.
Rev. Lett. \textbf{101}, 017003 (2008).

\bibitem{Shev2} S. N. Shevchenko, S. H. W. can der Ploeg, M. Grajcar, E.
Il'ichev, A. N. Omelyanchouk, and H.-G.Meyer, Phys. Rev. B. \textbf{78},
174527 (2008).

\bibitem{Oliver} W. D. Oliver, Y. Yu, J. C. Lee, K. K. Berggren, L.S.
Levitov, and T. P. Orlando, Science \textbf{310}, 1653 (2005).

\bibitem{Berns} D. M. Berns, W. D. Oliver, S. O. Valenzuela, A. V. Shytov,
K. K. Berggren, L.S. Levitov, and T. P. Orlando, Phys. Rev. Lett. \textbf{97}%
, 150502 (2006).

\bibitem{Berns2} D. M. Berns, M. S. Rudner, S. O. Valenzuela, K. K.
Berggren, W. D. Oliver, L.S. Levitov and T. P. Orlando, Nature \textbf{455},
51 (2008).

\bibitem{Rudner} M. S. Rudner, A. V. Shytov, L.S. Levitov, D. M. Berns, W.
D. Oliver, S. O. Valenzuela and T. P. Orlando, Phys. Rev. Lett. \textbf{101}%
, 190502 (2008).

\bibitem{Guozhu} G. Sun, X. Wen, Y. Wang, S. Cong, J. Chen, L. Kang, W. Xu,
Y. Yu, S. Han and P. Wu, Appl. Phys. Lett. \textbf{94}, 102502 (2009).

\bibitem{Guozhu2} G. Sun, J. Chen, Z. Ji, W. Xu, L. Kang, P. Wu, N. Dong, G.
Mao, Y. Yu and D. Xing, Appl. Phys. Lett. \textbf{89}, 082516 (2006).

\bibitem{Xueda} X. Wen, and Y. Yu, Phys. Rev. B. \textbf{79}, 094529 (2009).

\bibitem{Xueda2} X. Wen, Y. Wang and Y. Yu, arXiv:0912.0881.
\end{thebibliography}
\end{document}